%% file: uvmot_11_28_11_v8.tex
\documentclass[aps,prl,twocolumn,groupedaddress,showpacs]{revtex4-1}

\usepackage{graphicx}% Include figure files
\usepackage{dcolumn}% Align table columns on decimal point
\usepackage{bm}% bold math
\newcommand{\twos}[1]{\ensuremath{\hspace{-1pt}2\hspace{0.5pt}\hspace{0pt}S_{#1}\hspace{-1pt}}}
\newcommand{\twop}[1]{\ensuremath{\hspace{-1pt}2\hspace{0.5pt}\hspace{0pt}P_{#1}\hspace{-1pt}}}
\newcommand{\trep}[1]{\ensuremath{\hspace{-1pt}3\hspace{0.5pt}\hspace{0pt}P_{#1}\hspace{-1pt}}}

\newcommand{\fours}[1]{\ensuremath{\hspace{-1pt}4\hspace{0.5pt}\hspace{0pt}S_{#1}\hspace{-1pt}}}
\newcommand{\fivep}[1]{\ensuremath{\hspace{-1pt}5\hspace{0.5pt}\hspace{0pt}P_{#1}\hspace{-1pt}}}

\newcommand{\red}{\ensuremath{ \twos{1/2}\hspace{-0.0pt}\rightarrow\hspace{-0.0pt}\twop{3/2} }\ }
\newcommand{\uv}{\ensuremath{ \twos{1/2}\hspace{-0.0pt}\rightarrow\hspace{-0.0pt}\trep{3/2} }\ }
\newcommand{\uvk}{\ensuremath{ \fours{1/2}\hspace{-0.0pt}\rightarrow\hspace{-0.0pt}\fivep{3/2} }\ }

\newcommand{\TD}{\ensuremath{ T_{D} }}
\newcommand{\TR}{\ensuremath{ T_{R} }}

\newcommand{\one}{ \ensuremath{|1\rangle }\ }
\newcommand{\two}{ \ensuremath{|2\rangle }\ }

\newcommand{\isatred}{ \ensuremath{ I_{\text{sat}}^{2P}}  }
\newcommand{\isatuv}{  \ensuremath{ I_{\text{sat}}^{3P}}  }

\newcommand{\kb} { \ensuremath{k_{\mathrm{B}}}}

\begin{document}

\title{All-Optical Production of a Lithium Quantum Gas Using Narrow-Line Laser Cooling}

%\title{Highly Efficient All-Optical Method for Producing a Quantum Gas of $^{6}$Li}
%Demonstration of a $^{6}$Li magneto-optical trap using the \uv transition



\author{P. M. Duarte}
\author{R. A. Hart}
\author{J. M. Hitchcock}
  \altaffiliation[Present address: ]{Thermo Fisher Scientific, 2215 Grand Avenue Pkwy, Austin TX, 78728 USA}
\author{T. A. Corcovilos}
  \altaffiliation[Present address: ]{Department of Physics, The Pennsylvania State University, 104 Davey Laboratory, University Park, PA 16802 USA}
\author{T. -L. Yang}
\author{A. Reed}
  \altaffiliation[Present address: ]{Department of Physics, University of Colorado, 2000 Colorado Ave, Boulder, CO 80309 USA}
\author{R. G. Hulet}
\affiliation{Department of Physics and Astronomy and Rice Quantum Institute, Rice University, Houston, Texas 77005, USA}

\date{\today}

\begin{abstract}
We have used the narrow \uv transition in the ultraviolet (uv) to laser cool
and magneto-optically trap (MOT) $^6$Li atoms.  Laser cooling of lithium is
usually performed on the \red (D2) transition, and temperatures of $\sim$300
$\mu$K are typically achieved.  The linewidth of the uv transition is seven
times narrower than the D2 line, resulting in lower laser cooling temperatures.
We demonstrate that a MOT operating on the uv transition reaches temperatures
as low as 59 $\mu$K.  Furthermore, we find that the light shift of the uv
transition in an optical dipole trap at 1070 nm is small and blue-shifted,
facilitating efficient loading from the uv MOT.  Evaporative cooling of a two
spin-state mixture of $^6$Li in the optical trap produces a quantum degenerate
Fermi gas with $3 \times 10^{6}$ atoms a total cycle time of only 11~s.

\end{abstract}
\maketitle

The creation of quantum degenerate gases using all-optical
techniques~\cite{Chapman2001,Granade2002,*O'Hara2002,Weber2003,Jochim2003}
offers several advantages over methods employing magnetic traps.  Optical
potentials can trap any ground state, allowing selection of hyperfine sublevels
with favorable elastic and inelastic scattering properties.  In the case of
Fermi gases, the ability to trap atoms in more than one sublevel eliminates the
need for sympathetic cooling with another species~\cite{Truscott2001,
Shreck2001}, greatly simplifying the experimental setup.  All-optical methods
also facilitate rapid evaporative cooling since magnetically tunable Feshbach
resonances can be employed to achieve fast thermalization.

There are, however, challenges to all-optical methods.  An essential
prerequisite is an optical potential whose depth is sufficiently greater than
the temperature of the atoms being loaded.  The usual starting point is a laser
cooled atomic gas confined to a magneto-optical trap (MOT).  In a two-level
picture, atoms may be cooled to the Doppler limit, $\TD=\hbar\Gamma /(2\kb)$,
where $\Gamma/(2\pi)$ is the natural linewidth of the excited state of the
cooling transition~\cite{Hansch1975,Wineland1975}.  In many cases, however,
sub-Doppler temperatures can be realized due to the occurrence of polarization
gradient cooling arising from the multi-level character of real
atoms~\cite{Chu1998,*CohenTannoudji1998,*Phillips1998}.  Polarization gradient
cooling mechanisms are effective if the linewidth of the cooling transition is
small compared to the hyperfine splitting of the excited state, or if there is
a large degree of magnetic degeneracy in the ground state~\cite{XuPRL2003}.
The limit to cooling in these cases is the recoil temperature, $\TR=\hbar^2 k^2
/(2 m \kb )$, where $k$ is the wave number of the laser cooling transition and
$m$ is the mass of the atom.

\begin{figure} \includegraphics[width=0.48\textwidth]{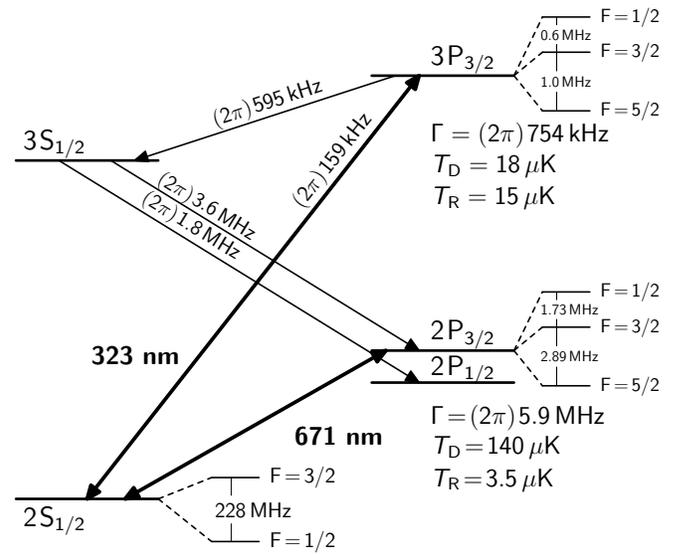}
\caption{\label{leveldiagram} Partial energy level diagram of $^6$Li showing relevant transitions. Lines in bold represent the
transitions used to laser cool atoms.  Lighter lines represent decay pathways
from the excited \trep{3/2} state. The corresponding decay rates are indicated~\cite{nistdatabase}.} \end{figure}

Polarization gradient cooling is found to be efficient for most of the alkali
metal atoms including Na, Rb, and Cs; MOTs of these species routinely attain
temperatures of $\sim$10 $\mu\text{K}$, which is not far above $\TR$.
Unfortunately, for Li and K, the elements most often employed in Fermi-gas
experiments, sub-Doppler cooling is ineffective in the presence of magnetic
fields, including those required for a MOT.  For Li, sub-Doppler cooling is
inhibited because the hyperfine splitting of the excited state is unresolved
(Fig.~\ref{leveldiagram}), thus limiting temperatures to $\sim$300~$\mu$K,
roughly twice the Doppler limit.  Sub-Doppler cooling has been attained in both
the bosonic~\cite{Landini2011,Gokhroo2011} and fermionic~\cite{Modugno1999}
isotopes of K by going from a MOT to a molasses phase, and appropriately ramping
the laser parameters~\cite{Landini2011,Gokhroo2011}.  While this can work at
high densities for the bosonic isotopes~\cite{Landini2011}, in the case of
fermionic $^{40}$K the lower temperatures are achieved at the expense of
density~\cite{Modugno1999}, which is not advantageous for loading and
evaporative cooling in a trap.    

%%%% FIGURE 1 : LEVELDIAGRAM %%%%

Laser cooling has also been demonstrated with atoms possessing ultranarrow,
dipole-forbidden transitions, such as the intercombination lines of the
alkaline earth atoms, where $\TD \lesssim \TR$.  These experiments produce
temperatures in the range of
1-10~$\mu\text{K}$~\cite{Katori1999,Binnewies2001,Curtis2001,Loftus2004}. In
this paper, we report a similar strategy for $^{6}$Li, where we form a MOT
using the narrow, yet still dipole-allowed, \uv transition shown in
Fig.~\ref{leveldiagram}.  The narrower linewidth of this transition compared to
the usual D2 line gives a correspondingly lower Doppler temperature.  Perhaps
just as significant for reaching high phase space densities as a prelude to
evaporative cooling, is that the shorter wavelength of this transition reduces
the optical depth and radiation trapping, which are important density limiting
effects~\cite{Sesko1991}.   We demonstrate the effectiveness of this laser
cooling scheme as the starting point for efficiently loading an optical dipole
trap and producing a degenerate Fermi gas with a large atom number.  The
advantage of using a uv transition to obtain high density was previously
demonstrated with metastable helium~\cite{Tychkov}. In that case, however, the
transition linewidth was identical with the conventionally used transition, and
the temperature was not reduced.  Also, the analogous narrow, but
dipole-allowed \uvk transition in K has recently been used to achieve lower
temperatures in a MOT~\cite{ThywissenarXiv}.

\begin{figure}
\includegraphics[width=0.48\textwidth]{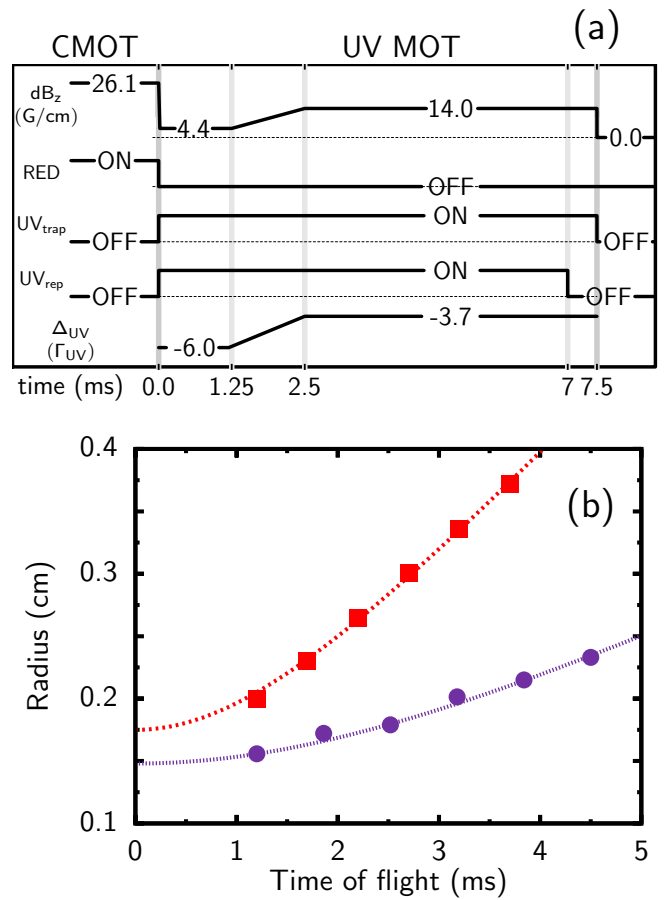}
\caption{\label{tofexpansion}
(Color online). 
(a) Timing diagram of the transfer sequence from
the CMOT to the uv MOT.
(b) Time-of-flight expansion of atoms released
from red and uv MOTs. The red squares (violet circles) represent the $1/e$ radius of
Gaussian fits to the spatial profile of freely expanding clouds of atoms
released from the CMOT (uv MOT).  The lines are fits to
ballistic expansions which give
temperatures of $ 288\,\mu\text{K}$ and $ 59\,\mu\text{K}$ and corresponding phase space densities of $2.5 \times 10^{-6}$ and $2.3 \times 10^{-5}$ for the CMOT and uv
MOT, respectively.  } \end{figure}

The experimental sequence begins by loading a conventional red MOT, operating
on the \red (D2) transition at $671\,\text{nm}$~\cite{Ritchie1995,Truscott2001,
Shreck2001}, from a laser slowed atomic beam.  Each of the six MOT beams
consists of light resonant with transitions between the $F=3/2$ and $F=1/2$
hyperfine ground states and the excited \twop{3/2} state, which we refer to as
the cooling and repumping beams, respectively.  These beams have $1/e^2$ radii
of $0.9\,\text{cm}$ and peak intensities per beam of $1.3\, \isatred$ on the
cooling transition and $0.4\, \isatred$ on the repumping transition,  where
$\isatred= 5.1\,\text{mW}/\text{cm}^2$ is the saturation intensity for the red
transition. After $\sim$5 seconds of loading, we collect $N\simeq 1.5 \times
10^{9}$ atoms at $T\simeq 1 \,\text{mK}$ and a peak density $n_0\simeq
2.4\times 10^{10} \,\text{cm}^{-3}$.  We then cool and compress the red MOT  by
simultaneously reducing the intensity and detuning of the cooling and repumping
light and increasing the magnetic field gradient.  At the end of this stage,
indicated as CMOT in Fig.\,\ref{tofexpansion}(a), the relevant quantities are
$N\simeq 1 \times 10^{9}$ atoms, $T\simeq 290 \,\mu\text{K}$, and $n_0\simeq
3.4 \times 10^{10} \,\text{cm}^{-3}$.

%%%% FIGURE 2 : TOFEXPANSION %%%%

The six beams for the uv MOT are overlapped with the red MOT beams using
dichroic mirrors.  The required uv light is generated by a frequency-doubled
laser system consisting of a grating stabilized diode laser operating at
$646\,\text{nm}$, a tapered amplifier, and an external ring doubling cavity
\footnote{The uv light source is a TA/DL-SHG 110 laser, manufactured by Toptica
Photonics}.  The output power at $323\,\text{nm}$ is limited to $
30\,\text{mW}$, but the total power incident at the atoms is only
$11\,\text{mW}$ due to power splitoff for frequency stabilization and to
losses from optics and vacuum viewports. The cooling (repumping) beams of the
uv MOT have $1/e^2$ radii of $0.33 \,\text{cm}$ ($0.40 \,\text{cm}$) and peak
intensities per beam of $0.2\,\isatuv$ ($0.02\,\isatuv$), where
$\isatuv=27.7\,\text{mW}/\text{cm}^2$ is the saturation intensity for the uv
transition.  Atoms are loaded into the uv MOT by abruptly reducing the magnetic
field gradient, turning off the red MOT light, and turning on the uv MOT light,
as shown in Fig.~\ref{tofexpansion}(a).  A small magnetic field gradient during
the $1.25 \,\text{ms}$ loading phase increases the effective capture volume,
and a large uv laser detuning helps capture high velocity atoms from the red
MOT.  The uv MOT captures $\sim$$5\times 10^8$ atoms, corresponding to an
efficiency of $50\%$.  The loss of atoms is most likely due to small uv beam
waists, which are limited by the total available power at $323\,\text{nm}$.
Following the loading phase the magnetic field gradient is increased linearly,
and the uv detuning is reduced over a period of $1.25 \,\text{ms}$. The values
of this steady-state uv MOT are chosen to optimize loading into the optical
dipole trap, as described below.  After a hold time of 5~ms, the temperature is
measured in time-of-flight by releasing and subsequently imaging the atoms by
pulsing on the red MOT cooling and repumping light at full intensity for 0.1 ms
while recording the fluorescence on a CCD camera.

Figure \ref{tofexpansion}(b) shows a comparison of the performance of the red
CMOT and the uv MOT. For the uv MOT, $T=59\,\mu\text{K}$ and $n_0\simeq 2.9
\times 10^{10} \,\text{cm}^{-3}$.  This corresponds to a phase space density
$\rho_{\mathrm{ps}}= n_0(h/\sqrt{2\pi m k_B T})^3= 2.3 \times 10^{-5}$, which
is an order of magnitude higher than that produced by the CMOT and thus
provides a significantly improved starting point for loading into an optical
dipole trap and subsequent evaporative cooling.  Lower temperatures may be
attained in the uv MOT but only at the cost of reducing the peak density.
Similarly, the density may be increased by using higher magnetic field
gradients and larger detunings of the uv cooling light, but this results in
higher $T$ and lower phase space density.  We also investigated a uv MOT in
which the repumping was done with the red transition.  In this case, the
temperature was the same as that obtained with uv repumping, but only half as
many atoms remained in the uv MOT and at a smaller density of $n_0\simeq 2.0
\times10^{10} \,\text{cm}^{-3}$.

%%%% FIGURE 3 : EVAPORATION %%%%
\begin{figure}
\includegraphics[width=0.48\textwidth]{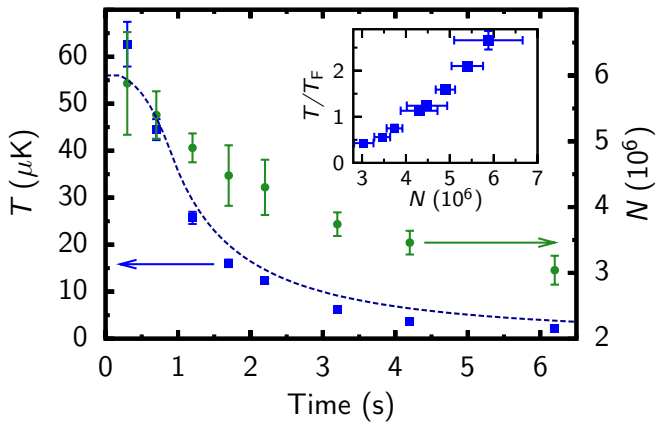}
\caption{\label{evaporation} 
(Color online). Number $N$ (green circles) and temperature $T$ (blue squares) of atoms in the optical trap during forced evaporation. Error bars for $N$ are one standard deviation of the mean value of $N$ for a sample of 5 measurements.  The dotted blue line is $U/5$, indicating the evaporation trajectory.  The inset shows $T/T_F$ for the points in the main plot.  For $T/T_F<1$ a surface fit to a polylog~\cite{Butts1997, Making2007} was used to determine $T/T_F$; otherwise, $T$ was measured by ballistic expansion and $T_F$ obtained from the mean value of $N$ and the measured trap frequencies.  Systematic uncertainties in all measured quantities are estimated to be 10\%.} \end{figure}

The light for the optical dipole trap is provided by a broadband fiber laser
operating at $1070\, \text{nm}$ with a nominal output power of $50 \,\text{W}$.
The trap consists of a single $38 \,\text{W}$ beam passed through the vacuum
chamber and then reintroduced to the chamber at an angle of $15\,^{\circ}$ with
respect to the first pass and with orthogonal polarization.  Each  beam is
cylindrically symmetric and focused to a $1/e^2$ radius of $73\,\mu\text{m}$ at
the point of intersection.  The trap depth per beam $U$ is $280\,\mu\text{K}$, and
the radial and axial frequencies of the trap are measured to be
$\omega_r=(2\pi) 3.8 \,\text{kHz}$ and $\omega_z=(2\pi) 475 \,\text{Hz}$,
respectively.

Atoms are loaded into the optical dipole trap by quickly turning on the trap
light when the steady-state values of the uv MOT have been reached, at 2.5\,ms
in the timing diagram of Fig. \ref{tofexpansion}(a).  We find that laser
cooling on the uv transition is effective in the trap, and that loading is
improved by leaving the uv MOT on for $5\,\text{ms}$ following turn on of the
trap.  The repumping light is turned off during the last $0.5\,\text{ms}$ of
loading, causing the atoms to be optically pumped into the $F=1/2$,
$m_F=\pm1/2$ hyperfine ground states, which we label as states \one and
\two\hspace{-4pt}.  The uv light and magnetic field gradient are then abruptly
switched off. A bias magnetic field is ramped within the next $20\,\text{ms}$
to $330\,\text{G}$, where the scattering length is $\sim$$-280 a_o$. We perform
evaporative cooling at $330\,\text{G}$ instead of near the Feshbach resonance
at $834\,\text{G}$ because we observe density dependent loss in the unitary
scattering regime that is fast enough to reduce the efficiency of evaporation~\cite{Du2009}.  This loss is unobservable at $330\,\text{G}$.

The evaporation trajectory is shown in Fig. \ref{evaporation}.  We leave the
dipole trap at full power for $200\,\text{ms}$ following loading to allow for
free evaporation of the atoms.    At this point $6 \times 10^6$ atoms remain at
$T=60\,\mu\text{K}$, corresponding to $T/T_{F}\approx2.7$, where
$T_F=\hbar\bar{\omega}(3 N)^{1/3}$ and
$\bar{\omega}=(\omega_{r}^2\omega_z)^{1/3}$.  With these parameters the peak
density is $n_0=2.7\times 10^{13} \,\text{cm}^{-3}$.  These are excellent
starting conditions for forced evaporation, which is performed by reducing the
trap laser power.  After $6 \,\text{s}$ of forced evaporation to $U =19
\,\mu\text{K}$, a degenerate sample is obtained with $3\times 10 ^6$ atoms at
$T/T_F=0.45$.

A key to the performance of this scheme is the ability of the uv MOT to achieve
both low laser cooled temperatures and high densities. The number of atoms
loaded into the dipole trap is predicted to depend exponentially on the ratio
of the depth of the dipole potential to the MOT temperature~\cite{OHara2001PRA}.  Although the optical dipole trap utilizes a high-power
laser, its depth is only $280\,\mu\text{K}$.  Such a depth would be ill-suited
to capture atoms from the red MOT.  In that case, either considerably greater
light power or a smaller trap volume would be needed.

%%%% FIGURE 4 : LIGHTSHIFT %%%%
\begin{figure}
\includegraphics[width=0.48\textwidth]{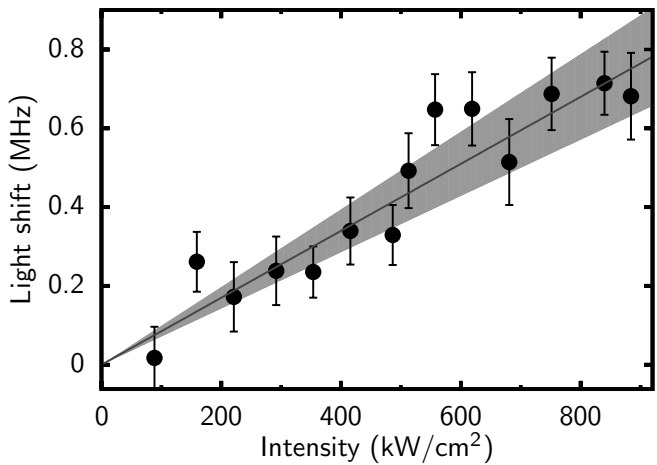}
\caption{\label{lightshift} Differential ac Stark shift of the \uv transition
as a function of intensity of the optical trapping light at
$\lambda=1070\,\text{nm}$.  The circles represent the center of a Gaussian fit of a loss
resonance (see text) and the error bars are 1 $\sigma$ statistical error of these fits.
The solid line is a linear fit to the resonance position with a slope of
$850(140) \,\mathrm{Hz/(kW/cm^{2})}$, where the uncertainty (shown by gray shading) represents the statistical uncertainty of the fit and a systematic uncertainty of 10\% on the value of the trap intensity.
} \end{figure}

Also intrinsic to the capability of the uv MOT to load atoms into the dipole
trap is the ability to continue to laser cool on the uv transition in the
optical trap, which is possible only if the differential AC Stark shift of the
\twos{1/2} and \trep{3/2} states produced by the dipole trap light is
sufficiently small~\cite{Katori2000,Grain2007}.  Otherwise, the light shift
would prevent uniform laser cooling in the trap, and depending on its sign,
could even cause heating.  Our dipole trap operates at $1070\, \text{nm}$ near
a predicted ``magic wavelength'' for the uv transition \footnote{M. Safronova,
private communication (2009).}, where the differential Stark shift vanishes. 

We measured the differential ac Stark shift of the uv cooling transition by
performing spectroscopy in the optical dipole trap.  The atoms are first cooled
to $T=3.5\,\mu\text{K}$ by $6\,\text{s}$ of forced evaporation.  Following
evaporation, the optical trap is adiabatically ramped to various peak
intensities, and the magnetic field is then quickly ramped to zero where the
atoms are illuminated by uv light with a frequency tuned near the \twos{1/2},
$F=1/2$, $\rightarrow\,$\trep{3/2} transition.  Resonant excitation causes
atoms to be optically pumped out of the $F=1/2$ ground state.  The population
remaining in \two is subsequently measured by absorption imaging at a field of
530 G.  Spectra are recorded at several trap intensities, and the center of
each is found by fitting to a Gaussian.  These resonance locations are
displayed as a function of peak intensity in Fig. \ref{lightshift}. We find
that at full trap depth, the uv transition is shifted to a 750 kHz greater
frequency than for free space.  This is consistent with our observation that
atoms are most efficiently loaded from the uv MOT when the uv laser frequency
is shifted to the blue by approximately one linewidth.  With this detuning the
temperature of atoms loaded into the optical trap is $70\,\mu\text{K}$, close
to the temperature in the MOT.  The fact that the light shift is small and to
the blue ensures that laser cooling is effective throughout the trap volume.

We have created a degenerate two-component Fermi gas with $3 \times 10^6$
$^{6}$Li atoms in 11 s using all-optical methods.  Our results demonstrate that
laser cooling on a narrow, but still dipole allowed, uv transition substantially
increases the atom number and production rate of a quantum degenerate gas.
Three features contribute to the success of this method: (1) the narrow
linewidth gives lower temperatures, enabling trapping with lower optical trap
depth, and hence, a larger trap volume for a given laser power; (2) the
differential light shift at the trapping wavelength is both small and to the
blue, which greatly enhances loading by permitting laser cooling to proceed in
the presence of the optical trap~\cite{Grain2007}; and (3) the short wavelength
cooling transition allows laser cooling to be effective even at higher densities.
Since $T_R \simeq T_D$ for the uv transition, the linewidth is sufficiently
broad to avoid the need for either a spectrally-broadened source or a
``quench'' laser to effectively broaden an ultra-narrow transition by coupling
it to a faster decaying excited state~\cite{Binnewies2001, Curtis2001,
Grain2007}.  Additionally, since $T_D$ for the D2 line is only 7 times larger
than $T_D$ for the uv transition, transfer from the red to uv MOT proceeds
efficiently without the need of an intermediate cooling
stage~\cite{Katori1999,Loftus2004}.  The generation of Fermi gases with large
number and densities is expected to be important in optical lattice experiments
that require unit filling, including the search for antiferromagnetic order in
the Hubbard model~\cite{Corcovilos2010}. 

\begin{acknowledgments}
We thank Tom Killian and Joseph Thywissen for useful discussions. We also acknowledge contributions
to the experiment by Kevin Jourde, Adrien Signoles, and Florian Emaury.  This
work was supported under ARO Grant No. W911NF-07-1-0464 with funds from the
DARPA OLE program, the AFOSR DURIP program, NSF, ONR, and the Welch Foundation
(Grant No. C-1133).  
\end{acknowledgments}

%\bibliography{uvmotbib}% Produces the bibliography via BibTeX.
\input{uvmot_11_28_11_v8.bbl}

\end{document}